\input amstex
\magnification=\magstep1 
\baselineskip=13pt
\documentstyle{amsppt}
\vsize=8.7truein \CenteredTagsOnSplits \NoRunningHeads
\def\UU{\Cal U}
\def\xx{\bold x}

\def\Pr{\bold{P\thinspace}}
\def\dist{\operatorname{dist}}
\def\EE{\bold{E\thinspace}}
\def\FF{\Cal F}
\def\GG{\Cal G}
\def\HH{\Cal H}

\topmatter
\title Computing the partition function of a polynomial on the Boolean cube \endtitle 
\author Alexander Barvinok  \endauthor
\address Department of Mathematics, University of Michigan, Ann Arbor,
MI 48109-1043, USA \endaddress
\email barvinok$\@$umich.edu  \endemail
\date May 2016 \enddate
\thanks  This research was partially supported by NSF Grant DMS 1361541.
\endthanks 
\keywords Boolean cube, polynomial, partition function, algorithm \endkeywords
\abstract For a polynomial $f: \{-1, 1\}^n \longrightarrow {\Bbb C}$, we define the partition function as the average of 
$e^{\lambda f(x)}$ over all points $x \in \{-1, 1\}^n$, where $\lambda \in {\Bbb C}$ is a parameter.
We present a quasi-polynomial algorithm, which, given such $f$, $\lambda$ and $\epsilon >0$ approximates the partition function within a relative error of $\epsilon$ in $N^{O(\ln n -\ln \epsilon)}$ time provided $|\lambda| \leq  (2L\sqrt{\deg f})^{-1}$, where $L=L(f)$ is a parameter bounding the Lipschitz constant of $f$ from above and $N$ is the number of monomials in $f$. As a corollary, we obtain a quasi-polynomial algorithm, which, given such an $f$ with coefficients $\pm 1$ and such that every variable enters not more than $4$ monomials, approximates the maximum of $f$ on $\{-1, 1\}^n$ within a factor of $O\left(\delta^{-1} \sqrt{\deg f} \right)$, provided the maximum is $N\delta$ for some $0<\delta  \leq 1$. If every variable enters not more than $k$ monomials for some fixed $k > 4$, we are able to establish a similar result when $\delta \geq (k-1)/k$.
\endabstract
\subjclass  90C09, 68C25, 68W25, 68R05 \endsubjclass

\endtopmatter

\document

\head 1. Introduction and main results \endhead

\subhead (1.1) Polynomials and partition functions \endsubhead
Let $\{-1, 1\}^n$ be the $n$-dimensional Boolean cube, that is, the set of all $2^n$ $n$-vectors 
$x=\left(\pm 1, \ldots, \pm 1\right)$ and let $f: \{-1, 1\}^n \longrightarrow {\Bbb C}$ be a polynomial with complex coefficients.
We assume that $f$ is defined as a linear combination of square-free monomials:
$$\split f(x)=\sum_{I \subset \{1, \ldots, n\}} \alpha_I & \xx^I \quad \text{where} \quad  \alpha_I \in {\Bbb C} \quad 
\text{for all} \quad I \quad \\ \text{and} \quad
&\xx^I =\prod_{i \in I} x_i \quad \text{for} \quad x=\left(x_1, \ldots, x_n\right), \endsplit \tag1.1.1$$
where we agree that $\xx^{\emptyset}=1$. As is known, the monomials $\xx^I$ for $I \subset \{1, \ldots, n\}$ constitute a basis of the vector space of functions $f: \{-1, 1\}^n \longrightarrow {\Bbb C}$.

We introduce two parameters measuring the complexity of the polynomial $f$ in (1.1.1). The {\it degree} of $f$ is the largest degree of a monomial $\xx^I$ appearing in (1.1.1) with a non-zero coefficient, that is, the maximum cardinality $|I|$ such that 
$\alpha_I \ne 0$:
$$\deg f=\max_{I:\ \alpha_I \ne 0} |I|.$$
We also introduce a parameter which controls the Lipschitz constant of $f$:
$$L(f)=\max_{i=1, \ldots, n} \sum \Sb I \subset \{1, \ldots, n\} \\ i \in I \endSb |\alpha_I|.$$ 
Indeed, if $\dist$ is the metric on the cube, 
$$\dist(x, y) = \sum_{i=1}^n |x_i - y_i| \quad \text{where} \quad x=\left(x_1, \ldots, x_n\right) \quad \text{and} \quad 
y=\left(y_1, \ldots, y_n \right)$$
then 
$$\left| f(x) - f(y)\right| \ \leq \ L(f) \dist (x, y). $$
We consider $\{-1, 1\}^n$ as a finite probability space endowed with the uniform measure. 

For $\lambda \in {\Bbb C}$ and a polynomial $f: \{-1, 1\}^n \longrightarrow {\Bbb C}$, we introduce the {\it partition function}
$$ {1 \over 2^n} \sum_{x \in \{-1, 1\}^n} e^{\lambda f(x)} = \EE e^{\lambda f}. $$

Our first main result bounds from below the distance from the zeros of the partition function to the origin.
\proclaim{(1.2) Theorem} Let $f: \{-1, 1\}^n \longrightarrow {\Bbb C}$ be a polynomial and let $\lambda \in {\Bbb C}$ be such that
$$|\lambda| \ \leq \ {0.55 \over L(f) \sqrt{\deg f}}.$$ 
Then
$$\EE e^{\lambda f} \ \ne \ 0.$$ 
If, additionally, the constant term of $f$ is $0$ then 
$$\left|\EE e^{\lambda f} \right| \ \geq \ (0.41)^n.$$
\endproclaim 

We prove Theorem 1.2 in Section 4.
As a simple example, let 
$f\left(x_1, \ldots, x_n\right)=x_1 + \ldots + x_n$. Then 
$$\EE e^{\lambda f} = \left(\EE e^{\lambda x_1} \right) \cdots \left(\EE e^{\lambda x_n}\right)= \left({e^{\lambda} + e^{-\lambda} \over 2} \right)^n.$$
We have $L(f)=\deg f=1$ and Theorem 1.2 predicts that $\EE e^{\lambda f} \ne 0$ provided $|\lambda| \leq 0.55$. Indeed, the smallest in the absolute value root of $\EE e^{\lambda f}$ is $\lambda = \pi i/2$ with $|\lambda| =\pi/2 \approx 1.57$. If we pick $f(x_1, \ldots, x_n)=a x_1 + \ldots + a x_n$ 
for some real constant $a >0$ then the smallest in the absolute value root of $\EE e^{\lambda f}$ is $\pi i/2a$ with $|\lambda|$ inversely proportional 
to $L(f)$, just as Theorem 1.2 predicts. It is not clear at the moment whether the dependence of the bound in Theorem 1.2 on $\deg f$ is optimal.

As we will see shortly, Theorem 1.2 implies that $\EE e^{\lambda f}$ can be efficiently computed if
$|\lambda|$ is strictly smaller than the bound in Theorem 1.2. When computing $\EE e^{\lambda f}$, we may assume that the constant term of $f$ is $0$, since 
$$\EE e^{\lambda(f+\alpha)}=e^{\lambda \alpha} \EE e^{\lambda f}$$
and hence adding a constant to $f$ results in multiplying the partition function by a constant. 

For a given $f$, we consider a univariate function
 $$\lambda \longmapsto \EE e^{\lambda f}.$$
As follows from Theorem 1.2, we can choose a branch of 
$$g(\lambda) = \ln \left(\EE e^{\lambda f}\right) \quad \text{for} \quad |\lambda| \ \leq \ {0.55 \over  L(f) \sqrt{\deg f}}$$
such that $g(0) = 0$.
It follows that $g(\lambda)$ is well-approximated by a low degree Taylor polynomial at $0$.

\proclaim{(1.3) Theorem} Let $f: \{-1, 1\}^n \longrightarrow {\Bbb C}$ be a polynomial with zero constant term and let
$$g(\lambda)=\ln \left(\EE e^{\lambda f}\right) \quad \text{for} \quad |\lambda| \ \leq \ {0.55 \over  L(f) \sqrt{\deg f}}.$$
For a positive integer $m \leq 5n$, let
$$T_m(f; \lambda)= \sum_{k=1}^m {\lambda^k \over k!} {d^k \over d\lambda^k} g(\lambda) \Big|_{\lambda=0}$$
be the degree $m$ Taylor polynomial of $g(\lambda)$ computed at $\lambda=0$. 
Then for $n \geq 2$
$$\left|g(\lambda) -T_m(f; \lambda)\right| \ \leq \  {50 n \over (m+1) (1.1)^m} +e^{-n}$$
provided 
$$ |\lambda| \ \leq \ {1 \over 2 L(f) \sqrt{\deg f}}.  \tag1.3.1$$
\endproclaim
In Section 3, we deduce Theorem 1.3 from Theorem 1.2.

As we discuss in Section 3.1, for a polynomial $f$ given by (1.1.1), the value of $T_m(f; \lambda)$ can be computed in $nN^{O(m)}$ time, where $N$ is the number of monomials in the representation (1.1.1).
 Theorem 1.3 implies that as long as $\epsilon \gg e^{-n}$, by choosing 
 $m=O\bigl(\ln n -\ln \epsilon \bigr)$, we can compute the value of $\EE e^{\lambda f}$ within relative error $\epsilon$ in $N^{O(\ln n - \ln \epsilon)}$ time provided $\lambda$ satisfies the inequality (1.3.1). For $\epsilon$ exponentially small in $n$, it is more efficient to evaluate $\EE e^{\lambda f}$ directly from the definition.
 
 \subhead (1.4) Relation to prior work \endsubhead This paper is a continuation of a series of papers by the author \cite{Ba15}, \cite{Ba16} and by the author and P. Sober\'on
\cite{BS14}, \cite{BS16} on algorithms to compute partition functions in combinatorics, see also \cite{Re15}. The main idea of the method is that the logarithm of the partition function is well-approximated by a low-degree Taylor polynomial at the temperatures above the phase transition (the role of the temperature is played by $1/\lambda$), while the phase transition is governed by the complex zeros of the partition function, cf. \cite{YL52}, \cite{LY52}.

The main work of the method consists of bounding the complex roots of the partition function, as in Theorem 1.2. While the general approach of this paper looks similar to the approach of \cite{Ba15}, \cite{Ba16}, \cite{BS14} and \cite{BS16} (a martingale type and a fixed point type arguments), in each case bounding complex roots requires some effort and new ideas. Once the roots are bounded, it is relatively straightforward to approximate the partition function as in Theorem 1.3.

Another approach to computing partition functions, also rooted in statistical physics, is the correlation decay approach, see \cite{We06} and \cite{BG08}.
While we did not pursue that approach, in our situation it could conceivably work as follows: given a polynomial $f: \{-1, 1\}^n \longrightarrow {\Bbb R}$ and a real $\lambda >0$, we consider the Boolean cube as a finite probability space, where the probability of a point $x \in \{-1, 1\}^n$ is $e^{\lambda f(x)}/\EE e^{\lambda f}$. This makes the coordinates $x_1, \ldots, x_n$ random variables. We consider a graph with vertices $x_1, \ldots, x_n$ and edges connecting two vertices $x_i$ and $x_j$ if there is a monomial of $f$ containing both $x_i$ and $x_j$. This introduces a graph metric on the variables $x_1, \ldots, x_n$ and one could hope that if $\lambda$ is sufficiently small, we have correlation decay: the random variable $x_i$ is almost independent on the random variables sufficiently distant from $x_i$ in the graph metric. This would allow us to efficiently approximate the probabilities 
$\Pr(x_i=1)$ and $\Pr(x_i=-1)$ and then recursively estimate $\EE e^{\lambda f}$. 

While both approaches treat the phase transition as a natural threshold for computability, the concepts of phase transition in our method (complex zeros of the partition function) and in the correlation decay approach (non-uniqueness of Gibbs measures) 
though definitely related and even equivalent for some spin systems \cite{DS87}, in general are different.

Theorem 1.3 together with the algorithm of Section 3.1 below implies that to approximate $\EE e^{\lambda f}$ within a relative error of $\epsilon >0$, it suffices to compute moments $\EE f^k$ for $k=O\left(\ln \epsilon^{-1}\right)$. This suggests some similarity with one of the results of \cite{K+96}, where (among other results) it is shown that the number of satisfying assignments of a DNF on $n$ Boolean variables is uniquely determined by the numbers of satisfying assignments for all possible conjunctions of
$k \leq 1+\log_2 n$ clauses of the DNF (though this is a purely existential result with no algorithm attached). Each conjunction of the DNF can be represented as a polynomial
$$\split \phi_j(x) =&{1 \over 2^{|S_j|}} \prod_{i \in S_j} (1+\epsilon_i x_i) \quad \text{where} \\ &S_j \subset \{1, \ldots, n\} \quad \text{and} \quad \epsilon_i \in \{-1, 1\}, \endsplit$$
and we let 
$$f(x)=\sum_{j=1}^m \phi_j(x).$$
 Then the number of points 
 $x \in \{-1, 1\}^n$ such that $f(x) >0$ is uniquely determined by various expectations 
 $\EE \phi_{j_1} \cdots  \phi_{j_k}$ for $k \leq 1 + \log_2 n$. The probability that $f(x) =0$ for a random point $x \in \{-1, 1\}^n$ sampled from the uniform distribution, can be approximated by $\EE e^{-\lambda f}$ for a sufficiently large $\lambda >0$. The expectations are precisely those that arise when we compute 
 the moments $\EE f^k$. It is not clear at the moment whether the results of this paper can produce an efficient way to compute the number of 
 satisfying assignments.

\head 2. Applications to optimization \endhead

\subhead (2.1) Maximizing a polynomial on the Boolean cube \endsubhead 
Let
 $f: \{-1, 1\}^n \longrightarrow {\Bbb R}$ be a polynomial with real coefficients defined by its monomial expansion (1.1.1). As is known, various computationally hard problems of discrete optimization, such as finding the maximum cardinality of an independent set in a graph, finding the minimum cardinality of a vertex cover in a hypergraph and the maximum constraint satisfaction problem can be reduced to finding the maximum of $f$ on the Boolean cube $\{-1, 1\}^n$, see, for example, 
\cite{BH02}. 
 
 The problem is straightforward if $\deg f \leq 1$. If 
 $\deg f=2$, it may already be quite hard even to solve approximately: Given an undirected simple graph $G=(V, E)$ with set 
$V=\{1, \ldots, n\}$ of vertices and set $E \subset {V \choose 2}$ of edges, one can express the largest cardinality of an 
{\it independent set} (a set vertices no two of which are connected by an edge of the graph), as the
maximum of 
$$f(x)={1 \over 2} \sum_{i=1}^n \left(x_i+1\right) - {1 \over 4} \sum_{\{i, j\} \in E} \left(1+x_i\right)\left(1+x_j\right)$$
on the cube $\{-1, 1\}^n$. It is an NP-hard problem to approximate the size of the largest independent set in a given graph on $n$ vertices within a factor of $n^{1-\epsilon}$ for any $0 < \epsilon \leq 1$, fixed in advance \cite{H\aa01}, \cite{Zu07}. If $\deg f=2$ and $f$ does not contain linear or constant terms, the problem reduces to the max cut problem in a weighted graph (with both positive and negative weights allowed on the edges), where there exists a polynomial time algorithm achieving an $O(\ln n)$ approximation factor, see \cite{KN12} for a survey.

If $\deg f \geq 3$, no efficient algorithm appears to be known that would outperform choosing a random point $x \in \{-1, 1\}^n$.
The maximum of a polynomial $f$ with $\deg f=3$ and no constant, linear or quadratic terms can be approximated within an $O\bigl(\sqrt{n/\ln n}\bigr)$ factor in polynomial time, see \cite{KN12}.
Finding the maximum of a general real polynomial (1.1.1) on the Boolean cube $\{-1, 1\}^n$ is equivalent to the problem of finding the maximum weight of a subset of a system of weighted linear equations over ${\Bbb Z}_2$  that can be simultaneously satisfied \cite{HV04}. Assuming that $\deg f$ is fixed in advance, $f$ contains $N$ monomials and the constant term of $f$ is $0$, a polynomial time algorithm approximating the maximum of $f$ within a factor of $O(\sqrt{N})$ is constructed in \cite{HV04}. More precisely, the algorithm from \cite{HV04} constructs a point $x$ such that $f(x)$ is 
within a factor of $O(\sqrt{N})$ from $\sum_I |\alpha_I|$ for $f$ defined by (1.1.1).
If $\deg f \geq 3$, it is unlikely that a polynomial time algorithm exists approximating the maximum of $f$ within a factor of 
$2^{{(\ln N)}^{1-\epsilon}}$ for any fixed $0< \epsilon \leq 1$ \cite{HV04}, see also \cite{H\aa01}.

  Let us choose
 $$\lambda={1 \over 2 L(f) \sqrt{\deg f}}$$
 as in Theorem 1.3. As is discussed in Section 3.5, by successive conditioning, we can compute in $N^{O(\ln n-\ln \epsilon)}$ time a point $y \in \{-1, 1\}^n$ which satisfies 
 $$e^{\lambda f(y)} \ \geq \ (1-\epsilon) \EE e^{\lambda f} \tag2.1.1$$
 for any given $0 < \epsilon \leq 1$.
 
 How well a point $y$ satisfying (2.1.1) approximates the maximum value of $f$ on the Boolean cube $\{-1, 1\}^n$? We consider polynomials with coefficients $-1$, $0$ and $1$, where the problem of finding an $x \in \{-1, 1\}^n$ maximizing $f(x)$ is equivalent to finding a vector in ${\Bbb Z}_2^n$ 
 satisfying the largest number of linear equations from a given list of linear equations over ${\Bbb Z}_2$.
 \proclaim{(2.2) Theorem} Let
  $$f(x) =\sum_{I \in \FF} \alpha_I \xx^I$$  
  be a polynomial with zero constant term, where $\FF$ is a family of non-empty subsets of the set 
  $\{1, \ldots, n\}$ and $\alpha_I =\pm 1$ for all $I \in \FF$. Let
  $$\max_{x \in \{-1, 1\}^n} f(x)= \delta |\FF| \quad \text{for some} \quad 0 \leq \delta \leq 1.$$
 Suppose further that every variable $x_i$ enters at most four monomials $\xx^I$ for $I \in \FF$.
 Then
 $$\EE e^{\lambda f} \ \geq \ \exp\left\{ {3 \lambda^2 \delta^2 \over 16}  |\FF| \right\} \quad \text{for} \quad 0 \leq \lambda \leq 1.$$
 \endproclaim

Since $\EE f=0$, the maximum of $f$ is positive unless $\FF =\emptyset$ and $f\equiv 0$.
It is not clear whether the restriction on the number of occurrences of variables in Theorem 2.2 is essential or an artifact of the proof. We 
can get a similar estimate for any number occurrences provided the maximum of $f$ is sufficiently close to $|\FF|$.

\proclaim{(2.3) Theorem} Let $$f(x) =\sum_{I \in \FF} \alpha_I \xx^I$$  
  be a polynomial with zero constant term, where $\FF$ is a family of non-empty subsets of the set 
  $\{1, \ldots, n\}$ and $\alpha_I =\pm 1$ for all $I \in \FF$. Let $k > 2$ be an integer and suppose that every variable $x_i$ enters at most $k$ monomials $\xx^I$ for $I \in \FF$.
  If
  $$\max_{x \in \{-1, 1\}^n} f(x) \ \geq \ {k-1 \over k}  |\FF|$$
then 
 $$\EE e^{\lambda f} \ \geq \ \exp\left\{ {3 \lambda^2 \over 16}  |\FF| \right\} \quad \text{for all} \quad 0 \leq \lambda \leq 1.$$
 \endproclaim

We prove Theorems 2.2 and 2.3 in Section 5. 
  
 Let $f$ be a polynomial of Theorem 2.2 and suppose that, additionally, $|I| \leq d$ for all $I \in \FF$, so that $\deg f \leq d$. We have  $L(f)\leq 4$ and we choose 
 $$\lambda={1 \over 8 \sqrt{d}}.$$
Let $y \in \{-1, 1\}^n$ be a point satisfying (2.1.1). Then 
$$f(y) \ \geq \ {1 \over \lambda} \ln \EE e^{\lambda f} + {\ln (1-\epsilon) \over \lambda} \ \geq \ {3 \lambda \delta^2  \over 16} |\FF| +
{\ln (1-\epsilon) \over \lambda}. $$
That is, if the maximum of $f$ is at least $\delta |\FF|$ for some $0 < \delta \leq 1$, we can approximate the maximum in quasi-polynomial time within a factor of $O\left(\delta^{-1} \sqrt{d}\right)$. Equivalently, if for some $0 < \delta \leq 0.5$ there is a vector in ${\Bbb Z}_2^n$ satisfying at least 
$(0.5 +\delta)|\FF|$ equations of a set $\FF$ of linear equations over ${\Bbb Z}_2$, where each variable enters at most 4 equations, in quasi-polynomial time we can compute a vector $v \in {\Bbb Z}_2^n$ satisfying at least $(0.5 + \delta_1)|\FF|$ linear equations from the system, where 
$\delta_1=\Omega(\delta^2/\sqrt{d})$ and $d$ is the largest number of variables per equation. 

Similarly, we can approximate in quasi-polynomial time the maximum of $f$ in Theorem 2.3 within a factor of $O(k\sqrt{d})$ provided the maximum is sufficiently close to $|\FF|$, that is, is at least
${k-1 \over k}|\FF|$.

In Theorems 2.2 and 2.3, one can check in polynomial time whether the maximum of $f$ is equal to $|\FF|$, as this reduces to testing the feasibility of a system of linear equations over ${\Bbb Z}_2$. However, for any fixed $0< \delta < 1$, testing whether the maximum is at least $\delta |\FF|$ 
is computationally hard, cf. \cite{H\aa01}.

H\aa stad \cite{H\aa 00} constructed a polynomial time algorithm that approximates the maximum of $f$ within a factor of $O(kd)$.
In \cite{B+15}, see also \cite{H\aa15}, a polynomial algorithm is constructed that finds the maximum of $f$ within a factor of 
$e^{O(d)} \sqrt{k}$, provided $f$ is an odd function. More precisely, the algorithm finds a point $x$ such that $f(x)$ is within a factor of
$e^{O(d)} \sqrt{k}$ from $|\FF|$.

\head 3. Computing the partition function \endhead

\subhead (3.1) Computing the Taylor polynomial of  $g(\lambda)=\ln \left( \EE e^{\lambda f}\right)$\endsubhead
First, we discuss how to compute the degree $m$ Taylor polynomial $T_m(f; \lambda)$ at $\lambda=0$ of the function
$$g(\lambda)= \ln \left( \EE e^{\lambda f}\right),$$
see Theorem 1.3.
Let us denote
$$h(\lambda) =\EE e^{\lambda f} \quad \text{and} \quad g(\lambda)=\ln h(\lambda).$$
Then 
$$g'={h' \over h} \quad \text{and hence} \quad h'=g' h.$$
Therefore,
$$h^{(k)}(0) = \sum_{j=1}^k {k-1 \choose j-1} g^{(j)}(0) h^{(k-j)}(0) \quad \text{for} \quad k=1, \ldots, m. \tag3.1.1$$
If we calculate the derivatives
$$h(0),\ h^{(1)}(0), \ldots, h^{(m)}(0), \tag3.1.2$$
then we can compute 
$$g(0),\ g^{(1)}(0), \ldots, g^{(m)}(0)$$
by solving a non-singular triangular system of linear equations (3.1.1) which has $h(0)=1$ on the diagonal.
Hence our goal is to calculate the derivatives (3.1.2).

We observe that 
$$h^{(k)}(0) = {1 \over 2^n} \sum_{x \in \{-1, 1\}^n} f^k(x) = \EE f^k. $$
For a polynomial $f$ defined by its monomial expansion (1.1.1) we have 
$$\EE f = \alpha_{\emptyset}.$$
We can consecutively compute the monomial expansion of $f, f^2, \ldots, f^m$ by using the following multiplication rule
for monomials on the Boolean cube $\{-1, 1\}^n$:
$${\bold x}^I {\bold x}^J = {\bold x}^{I \Delta J},$$
where $I \Delta J$ is the symmetric difference of subsets $I, J \subset \{1, \ldots, n\}$.
It follows then that for a polynomial $f: \{-1, 1\}^n \longrightarrow {\Bbb C}$ given by its monomial expansion (1.1.1) and 
a positive integer $m$, the Taylor polynomial
$$T_m(f; \lambda)=\sum_{k=1}^m {\lambda^k \over k!} {d^k \over d\lambda^k} g(\lambda)
 \Big|_{\lambda=0}$$
 can be computed in $nN^{O(m)}$ time, where $N$ is the number of monomials in $f$.

Our next goal is deduce Theorem 1.3 from Theorem 1.2.
 The proof is based on the following lemma.
\proclaim{(3.2) Lemma} Let $p: {\Bbb C} \longrightarrow {\Bbb C}$ be a univariate polynomial and suppose that 
for some $\beta > 0$ we have 
$$p(z) \ne 0 \quad \text{provided} \quad |z| \leq \beta.$$
Let $0 < \gamma < \beta$ and
for $|z| \leq \gamma$, let us choose a continuous branch of
$$g(z) =\ln p(z).$$
Let
$$T_m(z)=g(0) + \sum_{k=1}^m {z^k \over k!} {d^k  \over dz^k} g(z)  \Big|_{z=0}$$
be the degree $m$ Taylor polynomial of $g(z)$ computed at $z=0$.
Then for 
$$\tau={\beta \over \gamma} \ > \ 1$$ 
we have 
$$\left| g(z) - T_m(z)\right| \ \leq \ {\deg p \over (m+1)\tau^m (\tau-1)} \quad \text{for all} \quad |z| \leq \gamma.$$
\endproclaim 
\demo{Proof} Let $n=\deg p$ and let $\alpha_1, \ldots, \alpha_n$ be the roots of $p$,
so we may write 
$$p(z)=p(0) \prod_{i=1}^n \left(1 - {z \over \alpha_i}\right) \quad \text{where} \quad |\alpha_i| \geq \beta  \quad 
\text{for} \quad i=1, \ldots, n.$$
Then
$$g(z)=g(0)+ \sum_{i=1}^n \ln \left(1 - {z \over \alpha_i}\right),$$
where we choose the branch of the logarithm which is $0$ when $z=0$. Using the Taylor series expansion of the logarithm, we 
obtain
$$\ln \left(1 - {z \over \alpha_i}\right)=-\sum_{k=1}^m {z^k \over k \alpha_i^k}  +\zeta_m \quad \text{provided} \quad |z| \leq \gamma,$$
where
$$\left| \zeta_m \right| = \left| - \sum_{k=m+1}^{+\infty} {z^k \over k \alpha_i^k} \right| \ \leq \ 
\sum_{k=m+1}^{+\infty} {\gamma^k \over k \beta^k} \ \leq \ {1 \over (m+1)\tau^m (\tau -1)}.$$
Therefore,
$$g(z) = g(0) - \sum_{i=1}^n \sum_{k=1}^m {z^k \over k \alpha_i^k} +\eta_m \quad \text{for} \quad |z| \leq \gamma,$$
where
$$ \left| \eta_m \right| \ \leq \ {n \over (m+1) \tau^m (\tau -1)}.$$
It remains to notice that 
$$T_m(z)=g(0) - \sum_{i=1}^n \sum_{k=1}^m {z^k \over k \alpha_i^k}.$$
{\hfill \hfill \hfill} \qed
\enddemo

Next, we need a technical bound on the approximation of $e^z$ by its Taylor polynomial.
\proclaim{(3.3) Lemma} Let $\rho >0$ be a real number and let 
$m \geq 5 \rho$ be an integer.
Then 
$$\left| e^z - \sum_{k=0}^m {z^k \over k!} \right| \ \leq \ e^{-2\rho} \quad \text{for all} \quad z \in {\Bbb C}
\quad \text{such that} \quad |z| \leq \rho.$$
\endproclaim 
\demo{Proof} For all $z \in {\Bbb C}$ such that $|z| \leq \rho$, we have 
$$\split \left| e^z - \sum_{k=0}^m {z^k \over k!}  \right| = &\left| \sum_{k=m+1}^{+\infty} {z^k \over k!} \right| \ \leq \ \sum_{k=m+1}^{+\infty} {\rho^k \over k!} = {\rho^{m+1} \over (m+1)!} \sum_{k=0}^{+\infty} {\rho^k (m+1)! \over (k+m+1)!} \\  \leq \ &{\rho^{m+1} \over (m+1)!} \sum_{k=0}^{+\infty} {\rho^k \over k!} ={\rho^{m+1} e^{\rho} \over (m+1)!} \ \leq \ { \rho^{m+1} e^{\rho+m+1}\over (m+1)^{m+1}}.\endsplit$$
Since $m \geq 5\rho$, we obtain
$$\left| e^z - \sum_{k=0}^{+\infty} {z^k \over k!} \right| \ \leq \ { \rho^{m+1} e^{\rho+m+1} \over 5^{m+1} \rho^{m+1}} ={e^{\rho} \over (5/e)^{m+1}}\ \leq \ {e^{\rho} \over (5/e)^{5\rho}} \ \leq \ e^{-2\rho}.$$
and the proof follows.
{\hfill \hfill \hfill} \qed
\enddemo

\subhead (3.4) Proof of Theorem 1.3 \endsubhead Without loss of generality, we assume that $L(f)=1$. 
Since the constant term of $f$ is $0$, for any $x \in \{-1, 1\}^n$,  we have 
$$|f(x)| \ \leq \ \sum_{i=1}^n \sum_{I: \ i \in I} |\alpha_I| \ \leq \ n.$$
Applying Lemma 3.3, we conclude that
$$\left| e^{\lambda f(x)} - \sum_{k=0}^{5n} {\bigl(\lambda f(x)\bigr)^k \over k!} \right| \ \leq \ e^{-2n} \quad \text{for all} \quad 
x \in \{-1, 1\}^n \tag3.4.1$$
provided $|\lambda| \leq 1$.
Let 
$$p(\lambda)= 1+ \sum_{k=1}^{5n} {\lambda^k \over k!}  {d^k \over d \lambda^k}\left( \EE e^{\lambda f} \right) \Big|_{\lambda=0}$$
be the degree $5n$ Taylor polynomial of the function $\lambda \longmapsto \EE e^{\lambda f}$ at $\lambda=0$.
From (3.4.1) it follows that
$$\left| \EE e^{\lambda f} - p(\lambda) \right| \ \leq \ e^{-2n} \quad \text{provided} \quad |\lambda| \leq 1.$$
From Theorem 1.2, we conclude that 
$$p(\lambda) \ne 0 \quad \text{for all} \quad \lambda \in {\Bbb C} \quad \text{such that} \quad |\lambda| \ \leq \ 
{0.55 \over \sqrt{ \deg f}}$$
and, moreover,
$$\left| \ln p(\lambda) - \ln \left( \EE e^{\lambda f} \right) \right| \ \leq \ e^{-n} \quad \text{provided} \quad 
|\lambda| \ \leq \ {0.55 \over \sqrt{\deg f}} \quad \text{and} \quad n \geq 2. \tag3.4.2$$
Applying Lemma 3.2 with 
$$\beta={0.55 \over \sqrt{\deg f}}, \quad \gamma={0.5 \over \sqrt{\deg f}} \quad \text{and} \quad \tau={\beta \over \gamma}=1.1,$$
we conclude that for the Taylor polynomial of $\ln p(\lambda)$ at $\lambda=0$,
$$T_m(\lambda) = \ln p(0) + \sum_{k=1}^m {\lambda^k \over k!} {d^k \over d\lambda^k} \ln p(\lambda)\Big|_{\lambda=0}$$
we have 
$$\left| T_m(\lambda) - \ln p(\lambda) \right| \ \leq \ {50n \over (m+1) (1.1)^m} \quad \text{provided} \quad 
|\lambda| \ \leq \ {1 \over 2\sqrt{ \deg f}}. \tag3.4.3$$
It remains to notice that the Taylor polynomials of degree $m \leq 5n$ of the functions 
$$ \lambda \longmapsto \ln \left(\EE e^{\lambda f} \right) \quad \text{and} \quad  \lambda \longmapsto \ln p(\lambda)$$ at $\lambda=0$ coincide, since both are determined by the first $m$ derivatives 
of respectively $\EE e^{\lambda f}$ and $p(\lambda)$ at $\lambda=0$, cf. Section 3.1, and those derivatives coincide.
The proof now follows by (3.4.2) -- (3.4.3).
{\hfill \hfill \hfill} \qed

\subhead (3.5) Computing a point $y$ in the cube with a large value of $f(y)$ \endsubhead
We discuss how to compute a point $y \in \{-1, 1\}^n$ satisfying (2.1). We do it by successive conditioning and determine one coordinate of $y=\left(y_1, \ldots, y_n\right)$ at a time. Let $F^+$ and $F^-$ be the facets of the cube 
$\{-1, 1\}^n$ defined by the equations $x_n=1$ and $x_n=-1$ respectively for $x=\left(x_1, \ldots, x_n \right)$, $x \in \{-1, 1\}^n$. Then $F^+$ and $F^-$ can be identified with the $(n-1)$-dimensional cube $\{-1, 1\}^{n-1}$ and we have 
$$\EE e^{\lambda f}={1 \over 2} \EE \left( e^{\lambda f} | F^+ \right)  +{1 \over 2} \EE \left(e^{\lambda f} | F^-\right).$$
Moreover, for the restrictions $f^+$ and $f^-$ of $f$ onto $F^+$ and $F^-$ respectively, considered as polynomials on 
$\{-1, 1\}^{n-1}$, we have 
$$\deg f^+,\ \deg f^- \ \leq \ \deg f \quad \text{and} \quad L(f^+),\ L(f^-) \ \leq \ L(f).$$
Using the algorithm of Section 3.1 and Theorem 1.3, we compute $\EE\left(e^{\lambda f} | F^+\right)$ and 
$\EE \left(e^{\lambda f}| F^-\right)$ within a relative error $\epsilon/2n$, choose the facet with the larger computed value,
let $y_n=1$ if the value of $\EE\left(e^{\lambda f} | F^+\right)$ appears to be larger and let $y_n=-1$ if the value of 
$\EE\left(e^{\lambda f} | F^-\right)$ appears to be larger and 
proceed further by conditioning on the value of $y_{n-1}$. For polynomials with $N$ monomials, the complexity of the algorithm is $N^{O(\ln n)}$.

\head 4. Proof of Theorem 1.2 \endhead

To prove Theorem 1.2, we consider restrictions of the partition function onto faces of the cube. 
\subhead (4.1) Faces \endsubhead A {\it face} $F \subset \{-1, 1\}^n$ consists of the points $x$ where some of the coordinates of $x$ are fixed at $1$, some are fixed at $-1$ and others are allowed to vary (a face is always non-empty).
With a face $F$, we associate three subsets
$I_+(F), I_-(F), I(F) \subset \{1, \ldots, n \}$ as follows:
$$\split I_+(F)=&\bigl\{ i:\ x_i =1 \quad \text{for all} \quad x \in F,\ x=\left(x_1, \ldots, x_n \right)\bigr\}, \\
I_-(F)=&\bigl\{ i:\ x_i =-1 \quad \text{for all} \quad x \in F,\ x=\left(x_1, \ldots, x_n \right)\bigr\} \quad \text{and} \\
I(F)=&\{1, \ldots, n\} \setminus \left(I_+(F) \cup I_-(F)\right). \endsplit$$
Consequently,
$$\split F=\Bigl\{\left(x_1, \ldots, x_n\right) \quad \text{where} \quad &x_i =1 \quad \text{for} \quad i \in I_+(F) \quad \text{and}\\
&x_i=-1 \quad \text{for} \quad i \in I_-(F) \Bigr\}. \endsplit$$

In particular, if $I_+(F)=I_-(F)=\emptyset$ and hence $I(F)=\{1, \ldots, n\}$, we have $F=\{-1, 1\}^n$.
We call the number
$$\dim F =|I(F)|$$ 
the {\it dimension} of $F$.

For a subset $J \in \{1, \ldots,n\}$, we denote by $\{-1, 1\}^J$ the set of all points
$$x=\left(x_j: \ j \in J \right) \quad \text{where} \quad x_j = \pm 1.$$
Let $F \subset \{-1, 1\}^n$ be a face.
For a subset $J \subset I(F)$ and a point $\epsilon \in \{-1, 1\}^{J}$, $\epsilon=\left(\epsilon_j: \ j \in J \right)$, we define 
$$F^{\epsilon} =\bigl\{ x \in F,\ x=\left(x_1, \ldots, x_n\right):\ x_j=\epsilon_j \quad \text{for} \quad j \in J \bigr\}.$$
In words: $F^{\epsilon}$ is obtained from $F$ by fixing the coordinates from some set $J \subset I(F)$ of free coordinates to 
$1$ or to $-1$.
Hence $F^{\epsilon}$ is also a face of $\{-1, 1\}^n$ and we think of $F^{\epsilon} \subset F$ as a face of $F$. We can represent $F$ as a disjoint union
$$F= \bigcup_{\epsilon \in \{-1, 1\}^{J}} F^{\epsilon} \quad \text{for any} \quad J \subset I(F). \tag4.1.1$$

 \subhead (4.2) The space of polynomials \endsubhead Let us fix a positive integer $d$. We identify the set of all 
 polynomials $f$ as in (1.1.1) such that $\deg f \leq d$ and the constant term of $f$ is $0$ with ${\Bbb C}^N$, where
 $$N=N(n, d)=\sum_{k=1}^d {n \choose k}.$$
For $\delta > 0$, we consider a closed convex set $\UU(\delta) \subset {\Bbb C}^N$ consisting of the polynomials 
$ f: \{-1, 1\}^n \longrightarrow {\Bbb C}$ such that $\deg f \leq d$ and $L(f) \leq \delta$. In other words, $\UU(\delta)$ consists of the polynomials
 $$f(x)=\sum \Sb I \subset \{1, \ldots, n\} \\ 1 \leq |I| \leq d \endSb \alpha_I \xx^{I} \quad \text{where} \quad 
 \sum_{I: \ i \in I} |\alpha_I| \ \leq \ \delta \quad \text{for} \quad i=1, \ldots, n.$$

\subhead (4.3) Restriction of the partition function onto a face \endsubhead 
Let $f: \{-1, 1\}^n \longrightarrow {\Bbb C}$ be a polynomial and let $F \subset  \{-1, 1\}^n$ be a face. We define 
$$\EE\left(e^f | F \right)={1 \over 2^{\dim F}} \sum_{x \in F} e^{f(x)}.$$
We suppose that $f$ is defined by its monomial expansion as in (1.1.1) and consider $\EE\left(e^f | F \right)$ as a function of the coefficients 
$\alpha_I$. Using (4.1.1) we deduce
$$\split &{\partial \over \partial \alpha_J} \EE \left(e^f | F\right) = {1 \over 2^{\dim F}} \sum_{x \in F} \xx^J e^{f(x)} \\ &\quad=
{(-1)^{|I_-(F) \cap J|} \over 2^{|I(F)|}} \\ &\quad \times \sum \Sb \epsilon \in \{-1, 1\}^{I(F) \cap J} \\ \epsilon=(\epsilon_j: \ j \in I(F) \cap J) \endSb 
\left(\prod_{j \in I(F) \cap J} \epsilon_j\right) \sum_{x \in F^{\epsilon}} e^{f(x)} \\&\quad= {(-1)^{|I_-(F) \cap J|} \over 2^{|I(F) \cap J|}}\\
&\quad \times  \sum \Sb \epsilon \in \{-1, 1\}^{I(F) \cap J} \\ \epsilon=(\epsilon_j: \ j \in I(F) \cap J) \endSb 
\left(\prod_{j \in I(F) \cap J} \epsilon_j\right) \EE\left(e^f | F^{\epsilon} \right). \endsplit  \tag4.3.1$$

In what follows, we identify complex numbers with vectors in ${\Bbb R}^2 = {\Bbb C}$ and measure angles between non-zero complex numbers.

\proclaim{(4.4) Lemma} Let $0 < \tau \leq 1$ and $\delta >0$ be real numbers and 
let $F \subset \{-1, 1\}^n$ be a face. Suppose that for every $f \in \UU(\delta)$ we have $\EE\left(e^f | F\right) \ne 0$ and, moreover, 
for any $K \subset I(F)$ we have 
$$\left| \EE\left(e^f | F\right) \right| \ \geq \ \left({\tau \over 2}\right)^{|K|} \sum_{\epsilon \in \{-1, 1\}^K} 
\left| \EE\left(e^f, F^{\epsilon}\right)\right|.$$
Given $f \in \UU(\delta)$ and a subset $J \subset \{1, \ldots, n\}$ such that $|J| \leq d$,  let $\widehat{f} \in \UU(\delta)$ be the polynomial obtained from $f$ by changing the coefficient $\alpha_J$ of the monomial $\xx^J$ in $f$ to $-\alpha_J$ and leaving all other coefficients intact.
Then the angle between the two non-zero complex numbers $\EE\left(e^f | F\right)$ and $\EE\left(e^{\widehat{f}} | F\right)$ does not exceed
$${2 |\alpha_J| \over \tau^d}.$$
\endproclaim
\demo{Proof} Without loss of generality, we assume that $\alpha_J \ne 0$.

We note that for any $f \in \UU(\delta)$, we have $\widehat{f} \in \UU(\delta)$. Since $\EE\left(e^f | F\right) \ne 0$ for all $f \in \UU(\delta)$, we may consider a branch of $\ln \EE \left(e^f | F\right)$ for $f \in \UU(\delta)$.

Let us fix coefficients 
$\alpha_I$ for $I \ne J$ in 
$$f(x)=\sum\Sb I \subset \{1, \ldots, n\} \\ 1 \leq |I| \leq d \endSb \alpha_I \xx^{I} \tag4.4.1$$
and define a univariate function 
$$g(\alpha) = \ln \EE \left(e^f | F\right) \quad \text{where} \quad |\alpha| \leq |\alpha_J|$$
obtained by replacing $\alpha_J$ with $\alpha$ in (4.4.1).

 We obtain
$$g'(\alpha)={\partial \over \partial \alpha_J} \ln \EE \left(e^f | F \right) = 
\left({\partial \over \partial \alpha_J} \EE \left(e^f | F\right) \right) \Big/ \EE \left(e^f | F \right). \tag4.4.2$$
Let 
$$k=| I(F) \cap J| \ \leq \ |J| \ \leq \ d.$$
Using (4.3.1) we conclude that 
$$\left| {\partial \over \partial \alpha_J} \EE \left(e^f | F\right) \right| \ \leq \ {1 \over 2^{k}}\sum_{ \epsilon \in \{-1, 1\}^{I(F) \cap J} }
\left| \EE \left(e^f | F^{\epsilon}\right)\right|. \tag4.4.3$$
On the other hand, 
 $$\left|\EE\left(e^f | F\right)\right| \ \geq \ \left({\tau \over 2}\right)^k \sum_{ \epsilon \in \{-1, 1\}^{I(F) \cap J} }
\left| \EE\left(e^f | F^{\epsilon}\right)\right|. \tag4.4.4$$
Comparing (4.4.2) - (4.4.4), we conclude that
$$|g'(\alpha)| = \left| {\partial \over \partial \alpha_J} \ln \EE \left(e^f | F \right) \right| \ \leq \ {1 \over \tau^k} \ \leq \ {1 \over \tau^d}.$$
Then
$$\left| \ln \EE \left(e^f | F\right) -\ln \EE \left(e^{\widehat{f}} | F\right) \right| = \left| g\left(\alpha_J\right) - g\left(-\alpha_J\right) \right| \ \leq \
2 |\alpha_J|  \max_{|\alpha| \leq |\alpha_J| }  \left| g'(\alpha)\right|  \ \leq \ {2|\alpha_J| \over \tau^d} $$
and the proof follows.
{\hfill \hfill \hfill} \qed
\enddemo
 
\proclaim{(4.5) Lemma} Let $ \theta \geq 0 $ and $\delta > 0$ be real numbers such that $\theta \delta < \pi$, let $F \subseteq \{-1, 1\}^n$ be a face such that 
$\dim F < n$ and suppose that  $\EE \left(e^f | F\right) \ne 0$ for all $f \in \UU(\delta)$. Assume that for any $f \in \UU(\delta)$, for any $J \subset \{1, \ldots, n\}$ 
such that $|J| \leq d$, and for the polynomial $\widehat{f}$ obtained from $f$ by changing the coefficient 
$\alpha_J$ to $-\alpha_J$ and leaving all other coefficients intact, the angle between non-zero complex numbers 
$\EE\left(e^f | F\right)$ and $\EE\left(e^{\widehat{f}} | F\right)$ does not exceed $\theta |\alpha_J|$.
 
Suppose that $\widehat{F} \subset \{-1, 1\}^n$ is a face obtained from $F$ by changing the sign of one of the coordinates in $I_+(F) \cup I_-(F)$. Then $G=F \cup \widehat{F}$ is a face of $\{-1, 1\}^n$ and for 
 $$\tau=\cos {\theta \delta \over 2}$$
 we have 
 $$\left|\EE\left(e^f | G \right)\right| \ \geq \ {\tau \over 2} \left( \left| \EE \left(e^f | F\right) \right| + 
 \left| \EE \left(e^f | \widehat{F} \right) \right | \right)$$
 for any $f \in \UU(\delta)$.
\endproclaim 
\demo{Proof} Suppose that $\widehat{F}$ is obtained from $F$ by changing the sign of the $i$-th coordinate. Let $\tilde{f}$ be a polynomial obtained from $f$ by replacing the coefficients $\alpha_I$ by $-\alpha_I$ whenever $i \in I$ and leaving all other coefficients intact. Then $\tilde{f} \in \UU(\delta)$ and 
the angle between $\EE \left(e^f | F\right)$ and $\EE \left(e^{\tilde{f}} | F\right)$ does not exceed 
$$\theta \sum_{I:\ i \in I} |\alpha_I| \ \leq \ \theta \delta.$$
 On the other hand,
$\EE\left(e^{\tilde{f}} | F \right) =\EE \left(e^{f} | \widehat{F}\right)$ and 
$$\EE\left(e^f | G\right) ={1 \over 2}  \EE \left(e^f | F\right) + {1 \over 2} \EE\left(e^f | \widehat{F}\right) =
{1 \over 2} \EE\left(e^f | F\right) + {1 \over 2} \EE \left(e^{\tilde{f}} | F\right).$$
Thus $\EE\left(e^f | G\right)$ is the sum of two non-zero complex numbers, the angle between which does not exceed $\theta \delta < \pi$.
Interpreting the complex numbers as vectors in ${\Bbb R}^2 ={\Bbb C}$, we conclude that the length of the sum is at least as large as the length of the sum of the orthogonal projections of the vectors onto the bisector of the angle between them, and the proof follows.
{\hfill \hfill \hfill} \qed
\enddemo
 
\subhead (4.6) Proof of Theorem 1.2 \endsubhead
Let us denote $d=\deg f$.

 One can observe that the equation
 $${2  \over \cos \left(\dsize {\theta \beta \over 2}\right)}=\theta$$
 has a solution $\theta \geq 0$ for all sufficiently small $\beta > 0$. Numerical computations show that one 
 can choose 
 $$\beta = 0.55,$$
 in which case  
 $$\theta \approx 2.748136091.$$ Let 
 $$\delta ={\beta \over \sqrt{d}} = {0.55 \over \sqrt{d}}.$$
 We observe that 
 $$0 \ < \ \theta \delta  \ \leq \ \theta \beta \approx 1.511474850 \ < \ \pi. $$
Let
 $$\tau = \cos {\theta \delta \over 2}=\cos {\theta \beta \over 2\sqrt{d}}.$$
 In particular, 
 $$\tau \ \geq \ \cos {\theta \beta \over 2}  \approx 0.7277659962.$$
 Next, we will use the inequality 
 $$\left(\cos{\alpha \over \sqrt{d}}\right)^d \ \geq \ \cos \alpha \quad \text{for} \quad 0\leq \alpha \leq {\pi \over 2} \quad \text{and} \quad d \ \geq \ 1. \tag4.6.1$$
 One can obtain (4.6.1) as follows. Since $\tan(0)=0$ and the function $\tan \alpha$ is convex for $0 \leq \alpha < \pi/2$, we have 
 $$\sqrt{d} \tan {\alpha \over \sqrt{d}} \ \leq \  \tan \alpha \quad \text{for} \quad 0 \leq \alpha < {\pi \over 2}.$$
 Integrating, we obtain
 $$d \ln \cos {\alpha \over \sqrt{d}} \ \geq \ \ln \cos \alpha \quad \text{for} \quad 0 \leq \alpha < {\pi \over 2}$$
 and (4.6.1) follows.
 
Using (4.6.1), we obtain
$${2 \over \left(\cos {\dsize \theta \delta \over 2}\right)^d}={ 2 \over \left(\cos { \dsize \theta \beta \over 2 \sqrt{d}}\right)^d} \ \leq \ {2 \over \cos \left({\dsize \theta \beta  \over 2}\right)}= \theta. \tag4.6.2$$

 We prove by induction on $m=0, 1, \ldots, n$ the following three statements.
\bigskip
(4.6.3) Let $F \subset \{-1, 1\}^n$ be a face of dimension $m$. Then, for  any $f \in \UU(\delta)$, we have
$\EE\left(e^f | F\right) \ne 0$.
\medskip
(4.6.4) Let $F \subset \{-1, 1\}^n$ be a face of dimension $m$, let $f \in \UU(\delta)$ and let $\widehat{f}$ be a polynomial obtained from $f$ by changing one of the coefficients $\alpha_J$ to $-\alpha_J$ and leaving all other coefficients intact. Then the angle between two non-zero complex numbers 
$\EE\left(e^f | F\right)$ and $\EE\left(e^{\widehat{f}} | F\right)$ does not exceed $\theta |\alpha_J|$.
\medskip
(4.6.5) Let $F \subset \{-1, 1\}^n$ be a face of dimension $m$ and let $f \in \UU(\delta)$. Assuming that $m>0$ and hence $I(F) \ne \emptyset$, let us choose any $i \in I(F)$ and let $F^+$ and $F^-$ be the corresponding faces of $F$ obtained 
by fixing $x_i=1$ and $x_i=-1$ respectively. Then 
$$\left|\EE\left(e^f | F\right)\right| \ \geq\  {\tau \over 2} \left(\left| \EE \left(e^f |F^+ \right) \right| + \left|\EE \left(e^f | F^-\right) \right|\right).$$
\bigskip
If $m=0$ then $F$ consists of a single point $x \in \{-1, 1\}^n$, so
$$\EE\left(e^f | F \right)=e^{f(x)} \ne 0$$ 
and (4.6.3) holds.  Assuming that $\widehat{f}$ is obtained from $f$ by replacing the coefficient $\alpha_J$ with $-\alpha_J$ and leaving all other coefficients intact, we get 
$${\EE\left(e^f | F\right) \over \EE\left(e^{\widehat{f}} | F \right)} = \exp\left\{{2 \alpha_J \xx^J}\right\}.$$
Since 
$$|2 \alpha_J \xx^J| =2 |\alpha_J| \ \leq \  \theta |\alpha_J|,$$
the angle between $\EE\left(e^f | F\right)$ and $\EE\left(e^{\widehat{f}} | F\right)$ does not exceed $\theta |\alpha_J|$ and (4.6.4) follows.
The statement (4.6.5) is vacuous for $m=0$.

Suppose that (4.6.3) and (4.6.4) hold for faces of dimension $m< n$. Lemma 4.5 implies that if $F$ is a face of dimension $m+1$ and $F^+$ and $F^-$ are $m$-dimensional faces obtained by fixing $x_i$ for some $i \in I(F)$ to $x_i=1$ and $x_i=-1$ respectively, then
$$\split \left|\EE\left(e^f | F\right)\right| \ \geq \ & \left(\cos { \theta \delta \over 2 }\right) { \left|\EE\left(e^f | F^+\right) \right| +
\left|\EE\left(e^f | F^- \right)\right| \over 2} \\=
&{\tau \over 2} \left( \left|\EE\left(e^f | F^+\right)\right| +\left|\EE\left(e^f | F^-\right)\right| \right) \endsplit$$ and the statement
(4.6.5) holds for $(m+1)$-dimensional faces. 

The statement (4.6.5) for $(m+1)$-dimensional faces and the statement (4.6.3) for $m$-dimensional faces imply the statement 
(4.6.3) for $(m+1)$-dimensional faces.

Finally, suppose that the statements (4.6.3) and (4.6.5) hold for all faces of dimension at most $m+1$. Let us pick a face 
$F \subset \{-1, 1\}^n$ of dimension $m+1$, where $0 \leq m < n$.  Applying the condition of statement (4.6.5) recursively to the faces of $F$, we get that for any $K \subset I(F)$, 
$$\left|\EE\left(e^f | F\right)\right| \ \geq \ \left({\tau \over 2}\right)^{|K|} \sum_{\epsilon \in \{-1, 1\}^K} 
\left|\EE\left(e^f | F^{\epsilon}\right)\right|.$$
Then, by Lemma 4.4, the angle between two non-zero complex numbers $\EE\left(e^f | F\right)$ and
 $\EE\left(e^{\widehat{f}} | F\right)$ does not exceed
$${2 |\alpha_J| \over \tau^d}={2 |\alpha_J|  \over \left( \cos  {\dsize \theta \delta \over 2 } \right) ^d}\ \leq \ \theta |\alpha_J|$$
by (4.6.2), and the statement  (4.6.4) follows for faces of dimension $m+1$.

This proves that (4.6.3) -- (4.6.5) hold for faces $F$ of all dimensions. Iterating (4.6.5), we obtain that for any 
$f \in \UU(\delta)$, we have
$$\left|\EE e^f \right| \ \geq \ \left({\tau \over 2}\right)^n  \sum_{x \in \{-1, 1\}^n} |e^{f(x)}|.$$
Since for any $x \in \{-1, 1\}^n$ and for any $f \in \UU(\delta)$, we have 
$$|f(x)| \ \leq \ \sum_{i=1}^n \sum \Sb I \subset \{1, \ldots, n\} \\ i \in I \endSb |\alpha_I| \ \leq \ n \delta \ \leq \ \beta n, $$ 
we conclude that 
$$\left|\EE e^f \right| \ \geq \  \tau^n  e^{-\beta n} \ \geq \ (0.41)^n.$$
The proof follows since if $f: \{-1, 1\}^n \longrightarrow {\Bbb C}$ is a polynomial with zero constant term and 
$$|\lambda| \ \leq \ {0.55 \over L(f)\sqrt{\deg f}},$$
then $\lambda f \in \UU(\delta)$.
{\hfill \hfill \hfill} \qed

\head 5. Proofs of Theorems 2.2 and 2.3 \endhead

The proofs of Theorems 2.2 and 2.3 are based on the following lemma.

\proclaim{(5.1) Lemma} 
Let 
$$f(x) = \sum_{I \in \FF}  \alpha_I \xx^I$$
be a polynomial such that $\alpha_I \geq 0$ for all $I \in \FF$.
Then 
$$\EE e^f \ \geq \ \prod_{I \in \FF} \left({e^{\alpha_I} + e^{-\alpha_I} \over 2} \right).$$
\endproclaim
\demo{Proof}
Since 
$$e^{\alpha x} =\left({e^{\alpha} + e^{-\alpha} \over 2} \right) + x\left({e^{\alpha} - e^{-\alpha} \over 2} \right) \quad 
\text{for} \quad x =\pm 1,$$
we have 
$$\EE e^f = \EE \prod_{I \in \FF} e^{\alpha_I \xx^I} = \EE \prod_{I \in \FF} \left( \left( {e^{\alpha_I} +e^{-\alpha_I} \over 2} \right)
+ \xx^I \left({e^{\alpha_I} - e^{-\alpha_I} \over 2} \right) \right). \tag5.1.1$$
Since
$${e^{\alpha_I} - e^{-\alpha_I} \over 2} \geq 0 \quad \text{provided} \quad \alpha_I \geq 0$$
and 
$$\EE \left( \xx^{I_1} \cdots \xx^{I_k}\right) \ \geq \ 0 \quad \text{for all} \quad I_1, \ldots, I_k,$$
expanding the product in (5.1.1) and taking the expectation, we get the desired inequality.
{\hfill \hfill \hfill} \qed
\enddemo

Next, we prove a similar estimate for functions $f$ that allow some monomials with negative coefficients.

\proclaim{(5.2) Lemma} Let $f(x)=g(x)-h (x)$ where
$$ g(x)=\sum_{I \in \GG} \xx^{I}, \quad h(x) =\sum_{I \in \HH} \xx^{I}, \quad \GG \cap \HH =\emptyset.$$
Suppose that the constant terms of $g$ and $h$ are $0$ and that every variable $x_i$ enters not more than $k$ monomials of $f$ for some 
integer $k>0$.
Then 
$$\EE e^{\lambda f} \ \geq \ \exp\left\{{3 \lambda^2 \over 8} \left(|\GG|-(k-1)|\HH|\right)\right\} \quad \text{for} \quad 0 \leq \lambda \leq 1.$$
\endproclaim
\demo{Proof} Since $\EE f=0$, by Jensen's inequality we have 
$$\EE e^{\lambda f} \ \geq \ 1$$
and the estimate follows if $|\GG| \leq (k-1)|\HH|$. Hence we may assume that $|\GG| > (k-1)|\HH|$.

Given a function $f: \{-1, 1\}^n \longrightarrow {\Bbb R}$ and a set $J \subset \{1, \ldots, n\}$ of indices, we define a function (conditional expectation)
$f_J: \{-1, 1\}^{n-|J|} \longrightarrow {\Bbb R}$ obtained by averaging over variables $x_j$ with 
$j \in J$:
$$f_J\left(x_i: \ i \notin J\right) = {1 \over 2^{|J|}} \sum\Sb x_j = \pm 1 \\ j \in J \endSb f\left(x_1, \ldots, x_n\right).$$
In particular, $f_J=f$ if $J=\emptyset$ and $f_J=\EE f$ if $J=\{1, \ldots, n\}$.
We obtain the monomial expansion of $f_J$ by erasing all monomials of $f$ that contain $x_j$ with $j \in J$.
By Jensen's inequality we have 
$$\EE e^{\lambda f} \ \geq \ \EE e^{\lambda f_J} \quad \text{for all real} \quad \lambda. \tag5.2.1$$
Let us choose a set $J$ of indices with $|J| \leq |\HH|$ such that every monomial in $h(x)$ contains at least one variable $x_j$ with $j \in J$.
Then every variable $x_j$ with $j \in J$ is contained in at most $(k-1)$ monomials of $g(x)$ and hence 
$f_J$ is a sum of at least $|\GG| - (k-1)|\HH|$ monomials.

From (5.2.1) and Lemma 5.1, we obtain
$$\EE e^{\lambda f} \ \geq \ \EE e^{\lambda f_J} \ \geq \ \left({e^{\lambda} + e^{-\lambda} \over 2}\right)^{|\GG|-(k-1)|\HH|} 
\ \geq \ \left(1 + {\lambda^2 \over 2} \right)^{|\GG|-(k-1)|\HH|}. $$
Using that 
$$\ln (1+x) \ \geq \ x-{x^2 \over 2}=x\left(1-{x\over 2}\right) \quad \text{for} \quad x \geq 0, \tag5.2.2$$
we conclude that 
$$\EE e^{\lambda f} \ \geq \ \exp\left\{ {\lambda^2 \over 2} \left( 1- {\lambda^2 \over 4}\right) \left(|\GG|-(k-1)|\HH|\right)\right\}
\ \geq \ \exp\left\{ {3\lambda^2 \over 8} \left(|\GG|-(k-1)|\HH|\right)\right\} $$
as desired.
{\hfill \hfill \hfill} \qed
\enddemo

Now we are ready to prove Theorem 2.3.

\subhead (5.3) Proof of Theorem 2.3 \endsubhead Let $x_0 \in \{-1, 1\}^n$, $x_0=\left(\xi_1, \ldots, \xi_n\right)$ be a maximum point of $f$, so that 
$$\max_{x \in \{-1, 1\}^n} f(x)=f(x_0).$$
Let us define $\tilde{f}: \{-1, 1\}^n \longrightarrow {\Bbb R}$ by 
$$\tilde{f}\left(x_1, \ldots, x_n\right) = f\left(\xi_1 x_1, \ldots, \xi_n x_n \right).$$
Then 
$$\max_{x \in \{-1, 1\}^n} f(x) = \max_{x \in \{-1, 1\}^n} \tilde{f}(x), \quad \EE e^{\lambda f} = \EE e^{\lambda \tilde{f}}$$
and the maximum value of $\tilde{f}$ on the cube $\{-1, 1 \}^n$ is attained at $u=\left(1, \ldots, 1\right)$. Hence without loss of generality, we may assume that the maximum value of $f$ on the cube $\{-1, 1 \}^n$ is attained at $u=(1, \ldots, 1)$.

We write 
$$f(x)=g(x)-h(x) \quad \text{where} \quad g(x) = \sum_{I \in \GG} \xx^I \quad \text{and} \quad h(x) =\sum_{I \in \HH} \xx^I$$
for some disjoint sets $\GG$ and $\HH$ of indices. Moreover,
$$\max_{x \in \{-1, 1\}^n} f(x)=f(u) = |\GG| - |\HH| \ \geq \ {k-1 \over k} |\FF|.$$
Since 
$$|\GG| +|\HH| =|\FF|,$$
we conclude that 
$$|\GG| \ \geq \ {2k-1 \over 2k} |\FF| \quad \text{and} \quad |\HH| \ \leq \ {1 \over 2k} |\FF|.$$
By Lemma 5.2, 
$$\EE e^{\lambda f} \ \geq \ \exp\left\{ {3 \lambda^2 \over 8} \left(|\GG| - (k-1)|\HH| \right) \right\} \ \geq \ 
\exp\left\{ {3 \lambda^2 \over 16} |\FF| \right\}$$
as desired.
{\hfill \hfill \hfill} \qed

To prove Theorem 2.2, we need to handle negative terms with more care.

\proclaim{(5.4) Lemma} Let $f(x)=g(x)-h (x)$ where
$$ g(x)=\sum_{I \in \GG} \xx^{I}, \quad h(x) =\sum_{I \in \HH} \xx^{I}, \quad \GG \cap \HH =\emptyset$$
and 
$$|\GG| \ \geq \ |\HH|.$$
 Suppose that the constant terms of $g$ and $h$ are $0$ and that the supports $I \in \HH$ of monomials in $h(x)$ are pairwise disjoint. Then 
$$\EE e^{\lambda f} \ \geq \  \exp\left\{ {3 \lambda^2 \over 8} \left(\sqrt{|\GG|} - \sqrt{|\HH|} \right)^2 \right\} \quad \text{for} \quad 0 \leq \lambda \leq 1.$$
\endproclaim
\demo{Proof} By Jensen's inequality we have 
$$\EE e^{\lambda f} \ \geq \ \exp\left\{ \lambda \EE f \right\} =1,$$
which proves the lemma in the case when $|\GG| =|\HH|$. Hence we may assume that $|\GG| > |\HH|$. 

If $|\HH| =0$ then, applying Lemma 5.1, we obtain
$$\EE e^{\lambda f} =\EE e^{\lambda g} \ \geq \ \left({e^{\lambda} + e^{-\lambda} \over 2}\right)^{|\GG|} \ \geq \ \left(1 + {\lambda^2 \over 2}\right)^{|\GG|}.$$
Using (5.2.2), 
we conclude that 
$$\EE e^{\lambda f} \ \geq \ \exp\left\{{\lambda^2 \over 2}\left(1-{\lambda^2 \over  4}\right) |\GG| \right\} \ \geq \ \exp\left\{ {3\lambda^2 \over 8} |\GG|\right\},$$ 
which proves the lemma in the case when $|\HH|=0$. Hence we may assume that $|\GG| > |\HH| >0$.

Since the supports $I \in \HH$ of monomials in $h$ are pairwise disjoint, we have
$$\EE e^{\lambda h} = \prod_{I \in \HH} \EE e^{\lambda \xx^I} = \left( {e^{\lambda} + e^{-\lambda} \over 2}\right)^{|\HH|}. \tag5.4.1$$
Let us choose real $p, q\geq 1$, to be specified later, such that 
$${1 \over p} +{1 \over q} =1.$$
Applying the H\"older inequality, we get 
$$\EE e^{\lambda g/p} =\EE \left(e^{\lambda f/p} e^{\lambda h/p} \right) \ \leq \ \left(\EE e^{\lambda f}\right)^{1/p} 
\left( \EE e^{\lambda q h/p} \right)^{1/q}$$
and hence
$$\EE e^{\lambda f} \ \geq \ {\left( \EE e^{\lambda g/p} \right)^p \over \left(\EE e^{\lambda qh/p}\right)^{p/q}}.$$
Applying Lemma 5.1 and formula (5.4.1), we obtain
$$\EE e^{\lambda f} \ \geq \ \left({e^{\lambda/p} +e^{-\lambda/p} \over 2} \right)^{|\GG| p} \left( {e^{\lambda q/p} +e^{-\lambda q/p} \over 2} \right)^{-|\HH| p/q}.$$
Since
$$e^{x^2/2} \ \geq \ {e^x + e^{-x} \over 2} \ \geq \ 1+ {x^2 \over 2} \quad \text{for} \quad x \geq 0,$$
we obtain 
$$\EE e^{\lambda f} \ \geq\  \left(1 + {\lambda^2 \over 2 p^2}\right)^{|\GG| p} \exp\left\{-{\lambda^2 q |\HH| \over 2p} \right\}.$$
Applying (5.2.2),
we obtain
$$ \EE e^{\lambda f} \ \geq \ \exp\left\{ {\lambda^2  |\GG| \over 2 p} - {\lambda^2 q |\HH| \over 2p } -{\lambda^4 |\GG| \over 8 p^3}   \right\}.$$
Let us choose 
$$p={\sqrt{|\GG|} \over \sqrt{|\GG|} - \sqrt{|\HH|}} \quad \text{and} \quad q={\sqrt{|\GG|} \over \sqrt{|\HH|}}.$$
Then 
$$\split \EE e^{\lambda f} \ \geq \ &\exp\left\{ {\lambda^2 \over 2} \left( \sqrt{|\GG|} - \sqrt{|\HH|} \right)^2 -{\lambda^4 \left(\sqrt{|\GG|} - \sqrt{|\HH|}\right)^3 \over 8 \sqrt{|\GG|}}\right\} \\ =\ &\exp\left\{ {\lambda^2 \over 2} \left( \sqrt{|\GG|} - \sqrt{|\HH|} \right)^2 \left(1- {\lambda^2 \left(\sqrt{|\GG|} -\sqrt{|\HH|}\right)  \over 4 \sqrt{|\GG|}} \right)\right\} \\
\geq & \ \exp\left\{ {3 \lambda^2\over 8} \left(\sqrt{|\GG|} -\sqrt{|\HH|}\right)^2 \right\}\endsplit$$
and the proof follows.
{\hfill \hfill \hfill} \qed
\enddemo

\proclaim{(5.5) Lemma} Let $f(x)=g(x)-h (x)$ where
$$ g(x)=\sum_{I \in \GG} \xx^{I}, \quad h(x) =\sum_{I \in \HH} \xx^{I}, \quad \GG \cap \HH =\emptyset$$
and 
$$|\GG| \ \geq \ |\HH|.$$
Suppose that the constant terms of $g$ and $h$ are $0$, that every variable $x_i$ enters at most two monomials in $h(x)$ and that if $x_i$ enters exactly two monomials in $h(x)$ then $x_i$ enters at most two monomials in $g(x)$. Then for $0 \leq \lambda \leq 1$, we have 
$$\EE e^{\lambda f} \ \geq \  \exp\left\{ {3 \lambda^2 \over 8} \left(\sqrt{|\GG|} - \sqrt{|\HH|} \right)^2 \right\}.$$
\endproclaim
\demo{Proof} We proceed by induction on the number $k$ of variables $x_i$ that enter exactly two monomials in $h(x)$. If $k=0$ then the result follows by Lemma 5.4. 

Suppose that $k >0$ and that $x_i$ is a variable that enters exactly two monomials in $h(x)$ and hence at most two monomials in $g(x)$.
As in the proof of Lemma 5.2, let $f_i: \{0, 1\}^{n-1} \longrightarrow {\Bbb R}$ be the polynomial obtained from $f$ by averaging with respect to $x_i$.
As in the proof of Lemma 5.2, we have 
$$\EE e^{\lambda f} \ \geq \ \EE e^{\lambda f_i} \quad \text{where} \quad 
f_i(x) =\sum_{I \in \GG_i} \xx^I - \sum_{I \in \HH_i} \xx^I$$
and $\GG_i$, respectively $\HH_i$, is obtained from $\GG$, respectively $\HH$, by removing supports of monomials containing $x_i$. 
In particular,
$$|\HH_i| = |\HH| -2 \quad \text{and} \quad |\GG_i| \geq |\GG| -2.$$
Applying the induction hypothesis to $f_i$, we obtain
$$\split \EE e^{\lambda f} \ \geq \ &\EE e^{\lambda f_i } \ \geq \ \exp\left\{ {3 \lambda^2 \over 8} \left(\sqrt{|\GG_i|} - \sqrt{|\HH_i|} \right)^2 \right\}
\\ \geq\ &\exp\left\{ {3 \lambda^2 \over 8} \left(\sqrt{|\GG|-2} - \sqrt{|\HH|-2} \right)^2 \right\}
 \ \geq \ \exp\left\{ {3 \lambda^2 \over 8} \left(\sqrt{|\GG|} - \sqrt{|\HH|} \right)^2 \right\} \endsplit$$
and the proof follows.
{\hfill \hfill \hfill} \qed
\enddemo

Finally, we are ready to prove Theorem 2.2.

\subhead (5.6) Proof of Theorem 2.2 \endsubhead As in the proof of Theorem 2.3 of Section 5.3, without loss of generality we may assume that the maximum of $f$ is attained at $u=(1, \ldots, 1)$.

We write 
$$f(x) =g(x) -h(x) \quad \text{where} \quad g(x) = \sum_{I \in \GG} \xx^I \quad \text{and} \quad h(x)=\sum_{I \in \HH} \xx^I$$
for some disjoint sets $\GG$ and $\HH$ of indices. Moreover,
$$\max_{x \in \{-1, 1\}^n} f(x) =f(u) =|\GG| - |\HH| = \delta |\FF|.$$
Since 
$$|\GG| + |\HH| =|\FF|,$$
we conclude that 
$$|\GG| = {1+ \delta \over 2} |\FF| \quad \text{and} \quad |\HH| = {1-\delta \over 2} |\FF|. \tag5.6.1$$
For $i=1, \ldots, n$ let $\mu_i^+$ be the number of monomials in $g$ that contain variable $i$ and let $\mu_i^-$ be the number of monomials in $h$ that contain $x_i$. Then 
$$\mu_i^+ + \mu_i^- \ \leq \ 4 \quad \text{for} \quad i=1, \ldots, n. \tag5.6.2$$
If for some $i$ we have $\mu_i^+ < \mu_i^-$ then for the point $u_i$ obtained from $u$ by switching the sign of the $i$-th coordinate, we have 
$$f(u_i) = \left(|\GG| -2\mu_i^+\right) - \left(|\HH| - 2\mu_i^-\right) = |\GG|-|\HH| + 2\left(\mu_i^- - \mu_i^+\right) \ > \ f(u),$$
contradicting that $u$ is a maximum point of $f$. Therefore,
$$\mu_i^+ \ \geq \ \mu_i^- \quad \text{for} \quad i=1, \ldots, n$$ 
and, in view of (5.6.2), we conclude that 
$$\mu_i^- \ \leq \ 2 \quad \text{for} \quad i=1, \ldots, n \quad \text{and if} \quad \mu_i^-=2 \quad \text{then} \quad \mu_i^+=2.$$
By Lemma 5.5, 
$$\EE e^{\lambda f} \ \geq \ \exp\left\{ {3 \lambda^2 \over 8} \left( \sqrt{|\GG|} - \sqrt{|\HH|}\right)^2 \right\}.$$
Using (5.6.1), we deduce that 
$$\split \EE e^{\lambda f} \ \geq \  &\exp\left\{ {3 \lambda^2 \over 8} \left( \sqrt{1 +\delta \over 2} - \sqrt{1-\delta \over 2} \right)^2 |\FF| \right\}
\\ =\ &\exp\left\{ {3 \lambda^2 \over 8} \left( 1-\sqrt{1-\delta^2} \right) |\FF| \right\} \ \geq \ 
\exp\left\{ {3 \lambda^2 \delta^2 \over 16}  |\FF| \right\},\endsplit$$
which completes the proof.
{\hfill \hfill \hfill} \qed

\head Acknowledgments \endhead

I am grateful to Johan H\aa stad for advice and references on optimizing a polynomial on the Boolean cube and to the anonymous referees for careful reading of the paper and useful suggestions.

\enddocument
\end